\newif\ifcheckpagelimits
\checkpagelimitstrue
\checkpagelimitsfalse

\ifcheckpagelimits
 \documentclass[nofootinbib,prl,aps,twocolumn,showpacs,showkeys,%
 amsmath,amssymb,superscriptaddress,final,reprint,floatfix,longbibliography]{revtex4-1}
 \newcommand{\todo}[1]{}
\else
 \documentclass[prl,aps,twocolumn,showpacs,showkeys,%
 amsmath,amssymb,superscriptaddress,final,reprint,floatfix,longbibliography]{revtex4-1}
 \newcommand{\todo}[1]{{\pdfmargincomment[icon=Note,color=pink]{#1}}}
\fi

\usepackage{lineno}
  \usepackage{mathptmx}
\usepackage{subfigure}
\usepackage{dcolumn}
\usepackage{amsmath,amssymb}
\usepackage{bm}
\usepackage{color}
\usepackage{overpic}
\usepackage{latexsym}
\usepackage{epstopdf}
\usepackage{color}
\usepackage[english]{babel}
\usepackage{latexsym}
\usepackage{stmaryrd}

\usepackage{braket}

\definecolor{mygrey}{gray}{0.35}
\definecolor{myblue}{rgb}{0.2,0.2,0.8}
\definecolor{myzard}{cmyk}{0,0,0.05,0}
\definecolor{mywhite}{rgb}{1,1,1}
\definecolor{myred}{rgb}{1,0.,0.3}

\usepackage[colorlinks=true,citecolor=myblue,linkcolor=myred]{hyperref}

 \def\ee{\mathord{\rm e}}
 
 \def\ii{\mathord{\rm i}}

\def\half{\textstyle\frac{1}{2}}

\renewcommand{\ii}{{\rm i}}
\renewcommand{\ee}{{\rm e}}
\def\beq{\begin{equation}}
\def\eeq{\end{equation}}

\begin{document}

\title{Driven Spin-Boson Luttinger Liquids}

\author{Andreas Kurcz}
\affiliation{Instituto de F\'{\i}sica Fundamental, IFF-CSIC, Calle Serrano 113b, Madrid E-28006, Spain}
\author{Juan Jos\'e Garc\'{\i}a-Ripoll}
\affiliation{Instituto de F\'{\i}sica Fundamental, IFF-CSIC, Calle Serrano 113b, Madrid E-28006, Spain}
\author{Alejandro Bermudez}
\affiliation{Instituto de F\'{\i}sica Fundamental, IFF-CSIC, Calle Serrano 113b, Madrid E-28006, Spain}

\begin{abstract}
We introduce   a   lattice model   of interacting spins and bosons that leads to Luttinger-liquid physics, and allows for quantitative tests of the theory of bosonization by means of  trapped-ion or superconducting-circuit experiments. By using  a variational bosonization ansatz, we calculate the power-law decay of spin and boson correlation functions, and study their dependence on a single tunable parameter, namely a bosonic driving. For  small drivings,   Matrix-Product-States (MPS) numerical methods are shown to be efficient and validate our ansatz. Conversely,  even static MPS  become inefficient for large-driving regimes, such that the experiment  can potentially  outperform  classical numerics, achieving one of the  goals of quantum simulations.
\end{abstract}

\pacs{TBD}
\ifcheckpagelimits\else
\maketitle
\fi

Our understanding of matter  relies on simplified  models that try to capture the essence of experiments with limited microscopic control (e.g.  transport in solids). A  radically different 
approach is being pursued with
 ultracold atoms~\cite{QS_ultracold_atoms}, trapped ions~\cite{QS_trapped_ions}, and superconducting circuits~\cite{QS_cirquit_QED},  where current technology allows to design and test such models  microscopically. This  constitutes a new way of exploring paradigmatic, yet not fully-understood, quantum many-body problems~\cite{feynman,QS}. Besides,  this {\it synthetic quantum matter}   may also realise  exotic  models without condensed-matter counterparts that bring the possibility of testing  phenomena beyond  Landau's Fermi-liquid or symmetry-breaking theories~\cite{wen_book}.

The physics of one-dimensional (1D) strongly-correlated  models is a fertile ground for such fascinating phenomena~\cite{giamarchi_book}. Here,  the interplay of dimensionality and interactions  renders the Fermi-liquid concept of quasiparticles futile. As predicted by the theory of {\it Luttinger liquids} (LLs),  collective  bosons become the relevant excitations in   various fermionic, bosonic or spin models~\cite{phenom_bosonization}. Despite being 1D, LLs are not  mere theoretical artefacts, but  manifest themselves in   magnets~\cite{spinon_spectrum_magnets}, organic salts~\cite{LL_organic_salts}, carbon nanotubes~\cite{LL_carbon_nanotubes},  semiconducting wires~\cite{LL_semiconducting_quantum_wires}, and spin ladders~\cite{ll_xxz_magnetic_ladders}. However, with the exception of the latter, the limited control over the microscopic  parameters hampers a more quantitative test of LL-theory, where ab-initio predictions without adjustable parameters are confronted to  experiments~\cite{giamarchi_exp_review}. It is thus desirable to find new platforms where to assess the  universality  of  LLs  quantitatively.

Ultracold bosonic~\cite{cold_bosons_sine_Gordon_qpt_LL} and fermionic~\cite{cold_bosons_many_spins_LL_velocities} atoms are clear candidates for this quest, as they realise Hubbard-type interactions amenable of LL-theory~\cite{LL_cold_atoms}. Much less is known about 
trapped-ion and superconducting-circuit setups,  which  lead to various models with  spin-boson interactions~\cite{QS_trapped_ions,QS_cirquit_QED}. From a fundamental perspective, it would be interesting to study if such spin-boson synthetic matter  hosts a LL. Moreover, 
these devices would have the advantage that the scaling of any two-point correlator with  distance, which lies at the heart of LL-theory, can be  measured directly. In this work, we address this question, and show that certain driven spin-boson models yield a rich playground to test  LL-theory quantitatively.

{\it Spin-boson synthetic quantum matter.--} We consider a chain of lattice spacing $d$, hosting  $i\in\{1\dots N\}$ spins and bosons described by the  $\mathfrak{su}(2)$   $\{\sigma_i^{z},\sigma_i^{\pm}\}$, and bosonic $\{a_i^{\dagger},a_i^{\phantom{\dagger}}\}$ algebras. This system not only leads to various equilibrium phases, such as spin-boson Mott~\cite{jc_lattice} and Anderson~\cite{jc_lattice_disorder}  insulators, or spin-boson Ising  magnets~\cite{rabi_lattice}, but  also to  non-equilibrium phenomena, such as emerging causality~\cite{Lieb_robinson_spin_boson}. So far, LLs have not been explored in this context.

To fill in this gap, we  study a driven spin-boson  Hamiltonian $H=H_0+V$, with $H_0$ describing the uncoupled dynamics
\begin{equation}
H_0=\hspace{-.5ex}\sum_i-J_{\rm s}\sigma_{i+1}^+\sigma_i^--J_{\rm b}a_{i+1}^\dagger a_i^{\phantom{\dagger}}+\frac{\omega_{\rm b}}{2}a_i^\dagger a_i^{\phantom{\dagger}}-\Omega_{\rm d}\ee^{-\ii q_{\rm d}d i}a_i^{\phantom{\dagger}}+{\rm H.c.},
\label{eq:H0}
\end{equation}
where $J_{\rm s} (J_{\rm b})$ represents the spin (bosonic) tunnelling strengths,  $\omega_{\rm b}$ is the bosonic on-site energy ($\hbar=1$), and $\Omega_{\rm d}, q_{\rm d}$ are the driving amplitude and momentum, respectively. We also introduce a spin-boson interaction of strength $U$, namely
\begin{equation}
V=\sum_iU\sigma_i^za_i^\dagger a_i^{\phantom{\dagger}}.
\label{eq:V}
\end{equation}
Instances of this spin-boson model can be implemented with trapped ions or superconducting circuits, but we shall postpone the experimental details,  focusing first on its physical content. In Eq.~\eqref{eq:H0}, the spin part describes the well-known isotropic XY model, whose groundstate corresponds to a  half-filled Fermi sea of spinless fermions~\cite{XY_model}. In the absence of driving, the bosons minimize their energy in the global vacuum $\omega_{\rm b}>2J_{\rm b}$, such that the spin-boson interaction~\eqref{eq:V}  does not induce any  spin-boson LL behaviour. As shown below, as the driving is switched on, the bosonic mode of momentum $q_{\rm d}$ gets macroscopically populated. As a consequence, the spin-boson interaction  induces non-trivial correlations, which will be shown to correspond to a new phase: a  {\it spin-boson LL}.

{\it Variational ansatz.--} According to the above discussion, we introduce   spinless  fermions through a Jordan-Wigner mapping
$
\sigma_i^z=2c_i^{\dagger}c_i-1$, $\sigma_i^+=c_i^{\dagger}{\rm exp}({\ii\pi\sum_{j<i}c_j^{\dagger}c_j^{\phantom{\dagger}}})=(\sigma_i^-)^{\dagger}$~\cite{jordan_wigner}, and  build our  ansatz gradually. First,  we note that the macroscopic population of the driven  mode resembles the  Bogoliubov  theory    of superfluids, where  $a_q=\sum_i\ee^{-\ii qd i}a_i/\sqrt{N}\to\sqrt{N_{\rm d}}\delta_{q,q_{\rm d}}+\delta a_q,$ such that  $\delta a_q$ are  the Gaussian fluctuations for quasi-momentum $q\in[-\pi/d,\pi/d)$, and $\delta_{q,q_{\rm d}}$ is the Kronecker delta fixing the mode with macroscopic occupation $N_{\rm d}$~\cite{bogoliubov_weakly_interacting}. This can be accounted variationally  by the ansatz 
 \begin{equation}
 \nonumber
 \ket{\Psi_{\rm G}(\{\alpha_q,{\psi}_{\rm f}\})}=U_{\rm B}^\dagger \ket{\tilde{\Psi}_{\rm G}(\{{\psi}_{\rm f}\})},\hspace{1ex}U_{\rm B}=\ee^{-\sum_{q}(\alpha_qa_q^{\dagger}-{\rm H.c.})},
 %\label{eq:ansatz_1}
 \end{equation}
 where $\{\alpha_q,{\psi}_{\rm f}\}$ are the variational parameters, and $\ket{\tilde{\Psi}_{\rm G}}$ is a fiducial state.   Second, in order to account for the boson-fermion correlations caused by  the interaction~\eqref{eq:V}, we use a polaron  transformation~\cite{lang_firsov} for the fiducial state
  \begin{equation}
  \nonumber
  \ket{\tilde{\Psi}_{\rm G}(\{{\psi}_{\rm f}\})}= U_{\rm P}^{\dagger}\ket{0_{a}}\otimes\ket{{\psi}_{\rm f}},\hspace{1ex} U_{\rm P}=\ee^{\sum_{q,i}(f_{iq}a_q^{\dagger}-{\rm H.c.})c_i^\dagger c_i^{\phantom{\dagger}}},
  \end{equation}
  where $\{f_{iq}\}$ are such that the boson-fermion correlations are minimized in the unitarily-transformed picture,  $\ket{0_{a}}$ is the  boson vacuum, and  $\ket{{\psi}_{\rm f}}$ is a  general fermionic variational state.

The  variational minimization yields $\alpha_{ q}=\sqrt{N}(\Omega_{\rm d}/\omega_{ q})\delta_{q,q_{\rm d}}$  to lowest order in the interaction strength~\cite{supp_material}, where we have introduced the bosonic dispersion $\omega_q=\omega_{\rm b}-(U+2J_{\rm b}\cos qd)$.
This is the analogue of the macroscopic  population in the Bogoliubov  theory~\cite{bogoliubov_weakly_interacting}, which motivates neglecting contributions of   quartic terms. Within this approximation, one can see that the interaction~\eqref{eq:V} tries to create/annihilate Gaussian excitations conditioned on the fermion number $n_i=c_i^\dagger c_i^{\phantom{\dagger}}$ through a Holstein-type coupling~\cite{holstein}. This will lead to second-order processes where distant fermions exchange   Gaussian bosons virtually, resulting in a fermion-fermion long-range interaction~\cite{comment_polaron}. Additionally, as expected from the polaron transformation, a bosonic cloud will dress the  fermions  modifying their tunnelling (i.e. band narrowing).

To make this  description more quantitative, we specify the polaron parameters $f_{iq}=u_0\ee^{-\ii(q-q_{\rm d})di}/\omega_q\sqrt{N}$ with $u_0=2U\Omega_{\rm d}/\omega_{q_{\rm d}} $. Then, the variational minimization  sets  $\ket{{\psi}_{\rm f}}$ as the groundstate of the fermionic Hamiltonian \begin{equation}
H_{\rm f}=\sum_i(-\tilde{J}_{\rm s}c_i^\dagger c_{i+1}^{\phantom{\dagger}}+{\rm H.c.})+{\omega}_sn_i+\sum_{i}\sum_{j>i}
{V}_{i-j}n_in_j.
\label{long_range_XXZ}
\end{equation}
As announced above, we obtain   boson-mediated interactions  ${V}_{i-j}=-\sum_q\omega_q(f_{iq}f_{jq}^*+{\rm c.c.})$, and  renormalized tunnelings $\tilde{J}_{\rm s}=J_{s}\bra{0_{a}}\Pi_{q,q'} D_{a_q}(f_{iq})D_{a_{q'}}(-f_{i+1q'})\ket{0_{a}}$, where   the  generic displacement operator is $D_{a}(\alpha)={\rm exp}\big(\alpha a^{\dagger}-\alpha^* a\big)$. 

 To guarantee the  absence of negative frequencies and thus the stability of  bosons, we focus on $\lambda=2J_{\rm b}/(\omega_{\rm b}-U)<1$. By using  the displacement-operator algebra, the binomial theorem, and  some Taylor series~\cite{supp_material}, we find
\begin{equation}
\tilde{J}_{\rm s}={J}_{\rm s}\ee^{-\eta_1}, \hspace{1ex} {V}_{\ell}=V_0\cos\big(q_{\rm d}d\ell\big){\ee}^{-\frac{\ell d}{\xi_0}}, \hspace{1ex} \omega_{\rm s}=\frac{\Omega_{\rm d}u_0}{\omega_{q_{\rm d}}}+\frac{V_0}{2},
\label{eff_parameters}
\end{equation}
where the explicit dependences of  $\eta_1$, and $V_0, \xi_0$ on the experimentally tunable parameters  are listed in~\cite{explicit_tunn,explicit_int}.
 As announced, the dressed  tunnelling gets exponentially suppressed as the bosonic driving increases, and the interactions are long-ranged. For instance, letting $\omega_{\rm b}\approx 2J_{\rm b}$, the length scale $\xi_0$ diverges, such that the interaction  does not decay with  distance. Conversely, for  $\omega_{\rm b}\gg 2J_{\rm b}$,  interactions decay very rapidly. 
In contrast to the dressed tunnelling,  the  interaction strength  increases with the driving. Additionally,  its attractive/repulsive character oscillates along the chain, which corresponds  to frustration effects in the original spin model.

\begin{figure}
\centering
\includegraphics[width=1\columnwidth]{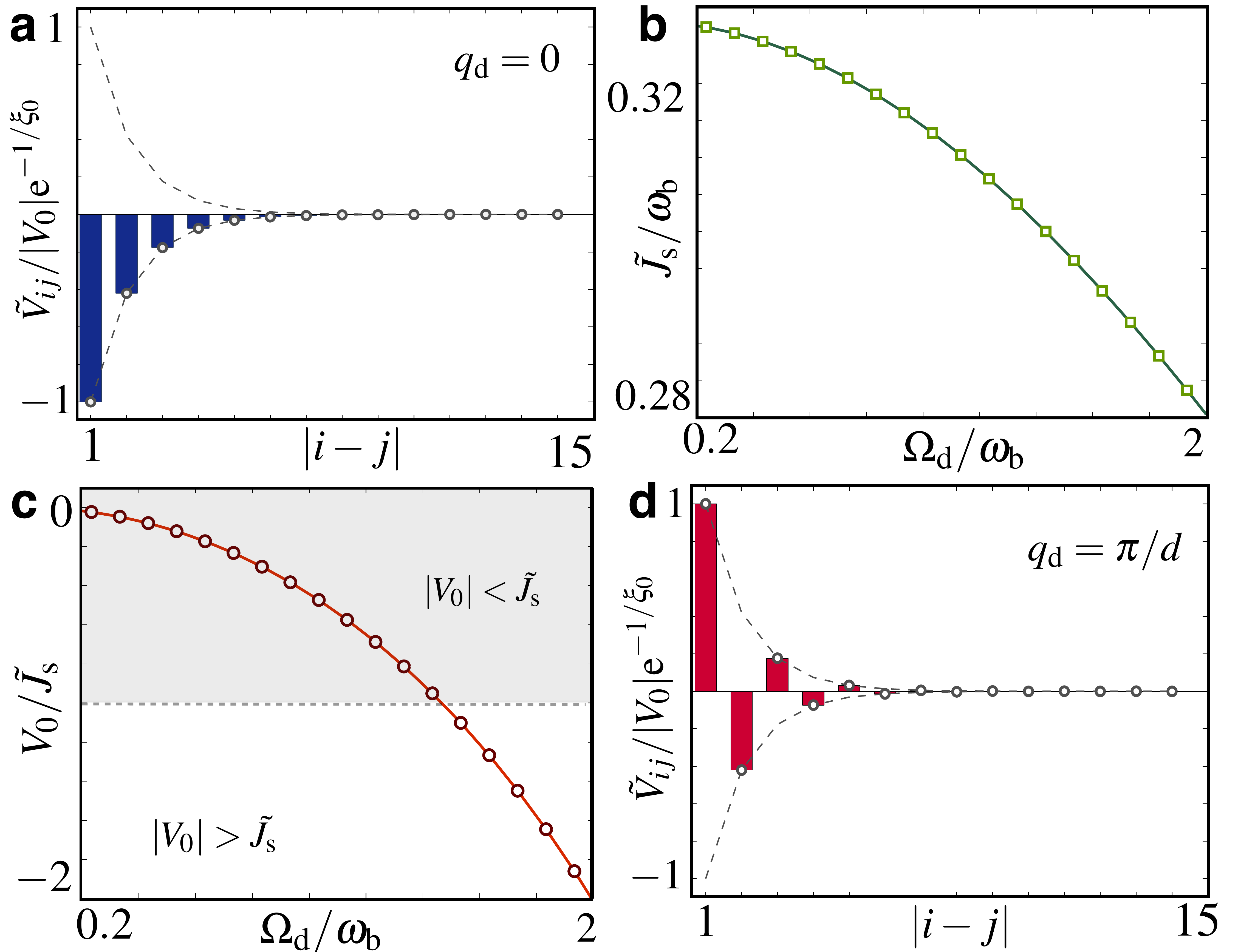}
\caption{ {\bf Parameters of the spinless fermion model: } {\bf (a)} Attractive  interactions in the  case $q_{\rm d}=0$, setting $d=1$,  $J_{\rm s}=J_{\rm b}=1=\omega_{\rm b}/3$, and $U=0.1$.  The bars (circles) represent the numerical (analytical~\eqref{eff_parameters}) result. {\bf (b)} Renormalised tunnelings, and {\bf (c) } ratio of the interaction  and tunnelling strengths, as a function of the driving. The solid line (symbols) corresponds to the numerics (analytics~\eqref{eff_parameters}). {\bf (d)}  Same as  {\bf (a)} but for  $q_{\rm d}=\pi/d$ leading to alternating interactions.}
\label{fig_parameters_XXZ}
\end{figure}

To test the correctness of Eq.~\eqref{eff_parameters}, we compare it to the corresponding exact expressions evaluated numerically. In  Figs.~\ref{fig_parameters_XXZ}{\bf (a)}-{\bf (b)}, we see that {\it (i)} the renormalization of the tunnelling and the exponential decay of the  interactions~\eqref{eff_parameters} are very accurate. Therefore, the interaction range can be tuned over  $\xi_0\in(0,\infty)$ by controlling $\lambda\in(0,1)$.  {\it (ii)} The dependence of the  degree of frustration on  the driving momentum~\eqref{eff_parameters} is also very accurate: while Fig.~\ref{fig_parameters_XXZ}{\bf (a)} corresponds to  unfrustrated attractive interactions, Fig.~\ref{fig_parameters_XXZ}{\bf (d)} shows that alternating attractive/repulsive interactions  occur for $q_{\rm d}=\pi/d$. Finally, {\it (iii)} the  ratio of the interactions to the  dressed tunnelling can be tuned across $|V_0|/\tilde{J}_{\rm s}=1$ by controlling a single parameter, the driving strength $\Omega_{\rm d}$ (Fig.~\ref{fig_parameters_XXZ}{\bf (c)}). This  is quite remarkable as we  started from the constraint $U\ll J_{\rm s},J_{\rm b}$ imposed by the Bogoliubov theory of the bosons. Nonetheless, the role of  fermion interactions is enhanced by increasing the driving strength.

%%%%%%%%%%
%%%%%%%%%%

{\it Bosonization and MPS.--} The nearest-neighbor limit of  the model~\eqref{long_range_XXZ}  is a paradigm of LLs~\cite{luther_peschel}. Thermodynamic quantities given by the  Bethe ansatz  can be combined with  LL-theory to obtain various correlation functions~\cite{haldane_ll}. Additionally, the numerical Density-Matrix-Renormalization-Group (DMRG)  gives  accurate predictions in this limit~\cite{dmrg,Heisenberg_dmrg}  used to benchmark the LL~\cite{xxz_nnn_dmrg}. The situation gets more involved for the full model~\eqref{long_range_XXZ}, since Bethe-ansatz integrability is lost, and  DMRG with long-range interactions is  more intricate (i.e. typically, models with only a few neighbors are studied~\cite{xxz_nnn_dmrg}). 

\begin{figure}
\centering
\includegraphics[width=1\columnwidth]{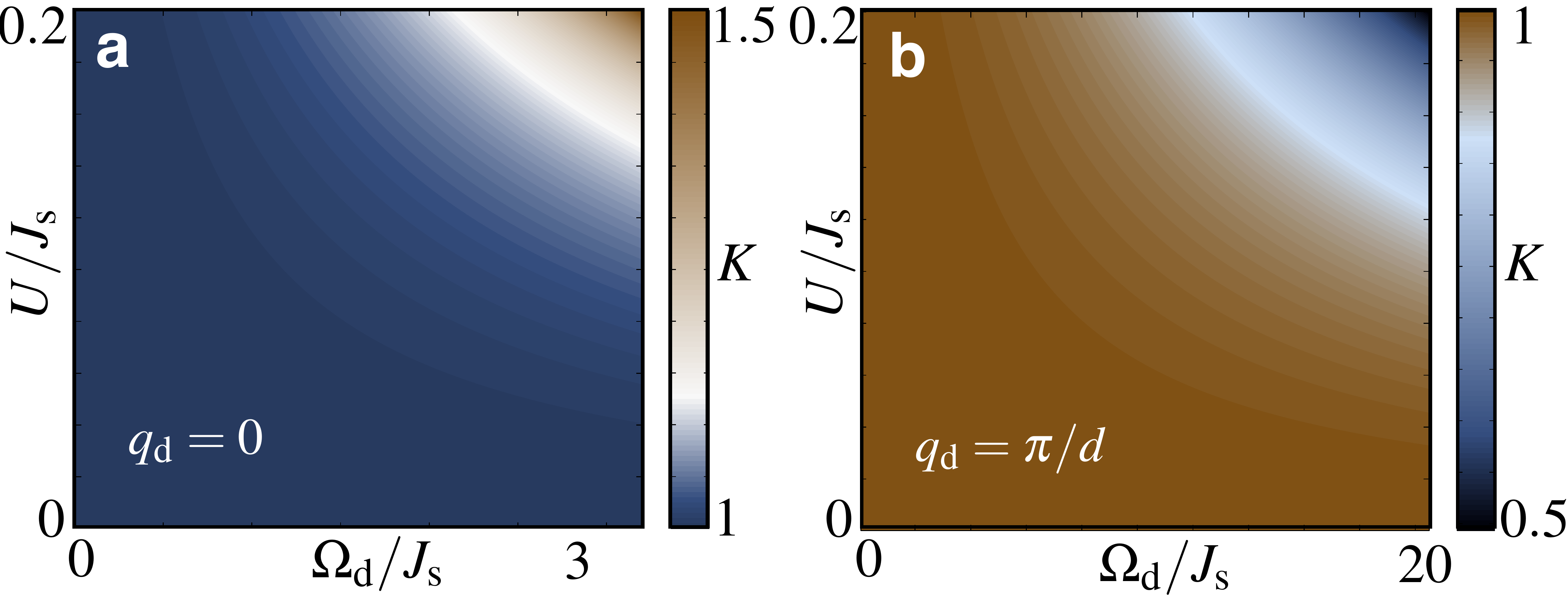}
\caption{ {\bf Luttinger parameter: } {\bf (a)} for the unmodulated case $q_{\rm d}=0$, and  {\bf (b)}  for  $q_{\rm d}=\pi$. Both figures  are calculated for $J_{\rm s}=J_{\rm b}=1=\omega_{\rm b}/3$, and different driving and interaction strengths.}
\label{fig_K_parameter}
\end{figure}

An analytical  LL-theory of the long-range model~\eqref{long_range_XXZ} can be obtained by   phenomenological bosonization~\cite{capponi}. We use instead the constructive approach~\cite{const_bosonization}, which allows us to find a constraint on the interaction range~\cite{supp_material}. Provided that {\it (i)}  $V_{i-j}\leq {\rm const}/|i-j|$  at large distances,  which is fulfilled by~\eqref{eff_parameters} except for $\lambda\to 1$, and {\it (ii)} $V_0\ll \tilde{J}_{\rm s}$, the low-energy excitations are described by two  bosonic branches
\beq
H_{\rm f}=\sum_{q>0}uq (d_{q{\rm +}}^\dagger d_{q{\rm +}}+d_{q{\rm -}}^\dagger d_{q{\rm -}}),
\label{bosonization}
\eeq   
characterized by the   sound velocity
\beq
u=2|\tilde{J}_{\rm s}|d\left(1+\frac{{V}_{0}}{2\pi|\tilde{J}_{\rm s}| }\frac{{\rm sinh}(\xi_0^{-1}d)\cos(q_{\rm d}d)}{{\rm cosh}^2(\xi_0^{-1}d)-\cos^2(q_{\rm d}d)}\right)^{1/2}.
\label{u_parameter}
\eeq
 The new  operators $d_{q\pm}$ are related to  particle-hole excitations of the original fermionic system by a squeezing transformation~\cite{supp_material} that depends on the Luttinger parameter
\beq
K=\left(1+\frac{{V}_{0}}{2\pi|\tilde{J}_{\rm s}| }\frac{{\rm sinh}(\xi_0^{-1}d)\cos(q_{\rm d}d)}{{\rm cosh}^2(\xi_0^{-1}d)-\cos^2(q_{\rm d}d)}\right)^{-1/2}.
\label{K_parameter}
\eeq 
In the absence of driving, we get $V_0=0$ and  recover the non-interacting value $K=1$. When we switch it on, it is possible to tune $K\lessgtr 1$ over a wide range of values as displayed in Fig.~\ref{fig_K_parameter}.

%%%%%%%%%%
%%%%%%%%%%

Following this discussion, the  fermionic part of our variational groundstate $\ket{\psi_{\rm f}}$ is the vacuum of the new bosonized modes $\ket{0_d}$, and we thus obtain a spin-boson LL groundstate 
 \beq
 \ket{\Psi_{\rm G}(\{\alpha_q,{\psi}_{\rm f}\})}=U_{\rm B}^\dagger U_{\rm P}^\dagger\ket{0_a}\otimes\ket{0_d}.
 \label{bosonized_var_gs}
 \eeq
We can now calculate any connected two-point correlator 
$
 C_{AB}(\ell)=\sum_i(\langle A_{\ell+i}B_i\rangle-\langle A_{\ell+i}\rangle\langle B_{i}\rangle)/N
 $ in the variational groundstate, provided that the operators $A,B$ are expressed in the bosonized picture. In this way, we can 
 test one of the distinctive features of LLs: the power-law decay of correlations.
 
 For instance, the diagonal spin correlators are
 \begin{equation}
 C_{\sigma^z\sigma^z}(\ell)=-\frac{2}{\pi^2}\left(\frac{K}{\ell^2}-(-1)^\ell\frac{1}{\ell^{2K}}\right),
 \label{szsz_LL}
\end{equation}
which coincide exactly with those of the  Heisenberg-Ising or XXZ model~\cite{luther_peschel}, albeit with a different Luttinger parameter~\eqref{K_parameter} due to the long range and possible frustration in~\eqref{long_range_XXZ}. The off-diagonal spin correlators are
\begin{equation}
  C_{\sigma^+\sigma^-}(\ell)=\frac{1}{2\pi}\frac{\ee^{-\eta_\ell}}{\ell^{\frac{1}{2K}}}\left(1-(-1)^\ell\frac{2}{\ell^{2K}}\right),
 \label{spsm_LL}
\end{equation}
which  display the power law of the bare XXZ  model~\cite{luther_peschel} with an additional  distance-dependent renormalization $\ee^{-\eta_\ell}$ due to the bosonic cloud  dressing the spins~\cite{explicit_tunn}. At long-distances, this polaron effect  is constant   $C_{\sigma^+\sigma^-}(\ell)\sim\ee^{-\eta_0/2}\ell^{-1/2K}$, and does not modify the  power-law exponent. 

The off-diagonal  bosonic correlators can also be understood from a polaron perspective
 \begin{equation}
 C_{a^{\dagger}a^{\phantom{\dagger}}}(\ell)=-\sum_{\tilde{\ell}}\frac{2}{\pi^2}\left(\frac{K}{\tilde{\ell}^2}-(-1)^{\tilde{\ell}}\frac{1}{\tilde{\ell}^{2K}}\right)C_{\ell,\tilde{\ell}}\hspace{0.3ex}\ee^{-\frac{|\ell-\tilde{\ell}|d}{\xi_0}},
 \label{ada_LL}
\end{equation}
which is a sum of diagonal spin correlations~\eqref{szsz_LL} with a polaron weight $C_{\ell,\tilde{\ell}}\hspace{0.3ex}{\rm exp}(-|\ell-\tilde{\ell}|d/\xi_0)$  exponentially suppressed at large  distances, where $C_{\ell,\tilde{\ell}}$ is listed in~\cite{explicit_cl}. For $\xi_0\gg d$, all terms except $\tilde{\ell}=\ell$ are negligible, and we thus find a power-law decay only determined by the Luttinger parameter.

\begin{figure}
\centering
\includegraphics[width=1\columnwidth]{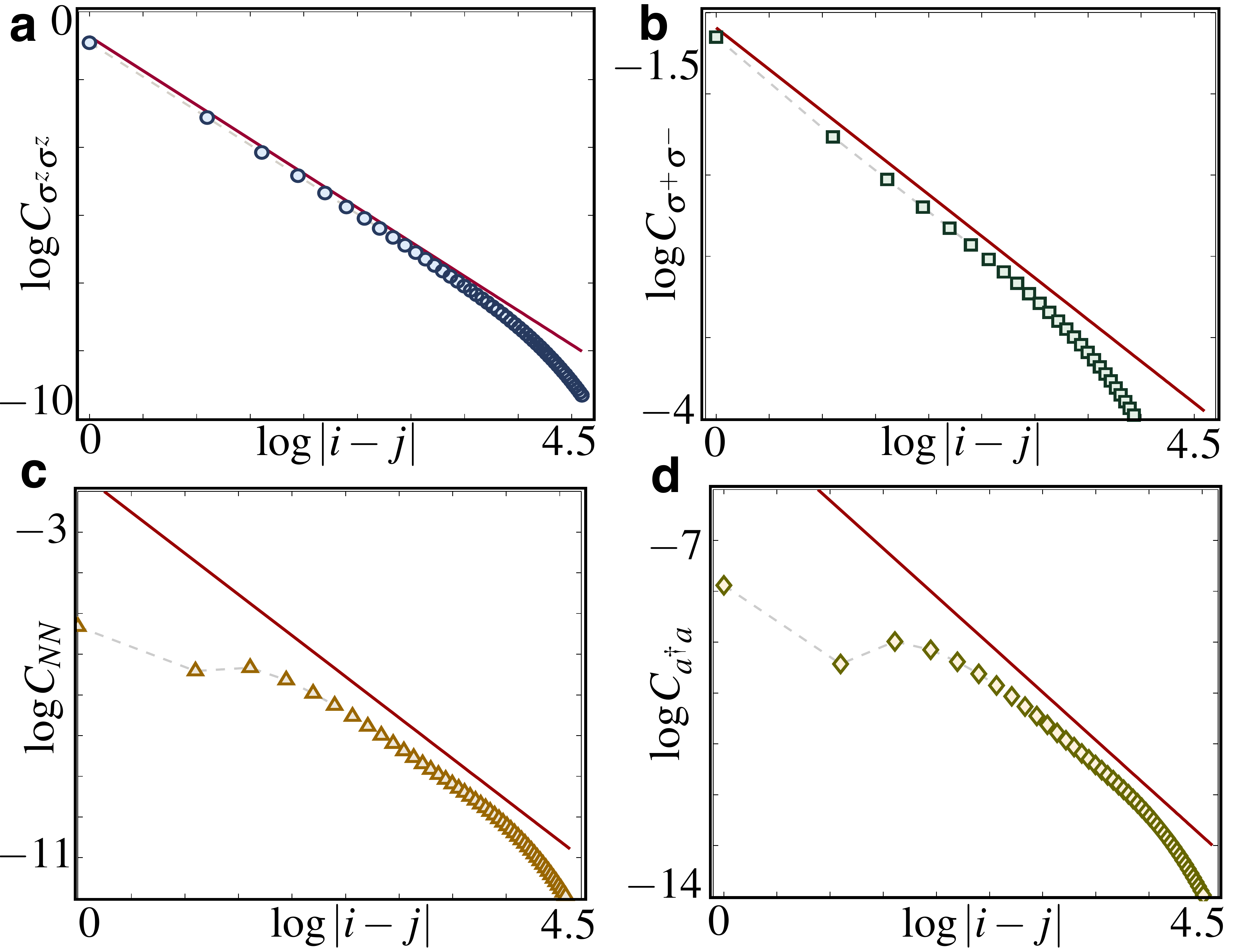}
\caption{ {\bf Two-point correlators:} Diagonal {\bf (a)},{\bf (c)} and off-diagonal {\bf (b)}, {\bf (d)} spin and boson correlators for $J_{\rm s}=J_{\rm b}=1=\omega_{\rm b}/3$,  $U=0.15$, and $\Omega_{\rm d}=1.5$ with $q_{\rm d}=0$. {\bf (c)}-{\bf (d)}. The symbols are numerical  data, and the red solid  lines represent the power-law decay obtained from fitting the LL-predictions~\eqref{szsz_LL}-\eqref{ada_LL} at  distances up to $|i-j|=100$. For $C_{NN}$, where $N=a^{\dagger} a$ , the analytical expression is only valid for $\xi_0\gg d$, where $C_{NN}(\ell)\approx (u_0/(\omega_{\rm b}-U))^4C_{\sigma^z\sigma^z}(\ell)$.}
\label{fig_correlations}
\end{figure}

Finally, we have  concrete  power-law predictions~\eqref{szsz_LL}-\eqref{ada_LL} that can be numerically benchmarked. We use DMRG-type methods based on matrix product states (MPS) for the thermodynamic limit~\cite{itebd}. By implementing the   original short-range spin-boson model~\eqref{eq:H0}-\eqref{eq:V}, we avoid the intricacies associated to  the long-range  model~\eqref{long_range_XXZ} mentioned above. 
The results displayed in Fig.~\ref{fig_correlations}  agree with the aforementioned power-law  decay at intermediate distances, and depart at longer distances due to technical limitations in the MPS dimension. These results  confirm the  validity of our ansatz. 

%%%%%%%%%%
%%%%%%%%%%

{\it Experimental LLs.--} Experiments on this versatile LL can access different regimes by controlling  a single parameter, the driving (Fig.~\ref{fig_K_parameter}). As a bonus,  regimes  that differ markedly from $K=1$ require very large drivings, which increases  both the bosonic population $|\alpha_{q_{\rm d}}|^2\propto\Omega_{\rm d}^2$ and the dimension of the MPS. Therefore,  MPS simulations,   limited by  available computing resources, will eventually cease to be trustworthy. Instead, the experiment  would act as a reliable quantum simulator~\cite{QS} capable of beating its classical MPS counterpart. Trapped-ion (TI) and superconducting-circuit (SC) architectures meet the requirements for this LL quantum simulator.

We first focus on the bosonic degrees of freedom. For laser-cooled  linear  chains of TIs~\cite{wineland_review}, the bosons are  the  transverse vibrational excitations of each ion, and  display Coulomb-induced dipolar tunnelings $J_{\rm b}a_i^{\dagger}a_j/|i-j|^3$, where $J_{\rm b}/2\pi\sim $1-10kHz. The  driving is due to an oscillating potential  at one of the electrodes, which has a frequency  $\omega_{\rm d}=\omega_{\rm t}+\Delta$ with detuning $\Delta$ from the transverse trap frequency $\omega_{\rm t}/2\pi\sim$1MHz. This  leads to $\Omega_{\rm d}/2\pi\sim$1-100kHz, and $q_{\rm d}=0$. To obtain   $q_{\rm d}\neq 0$, one should use instead the ac-Stark shift of a pair of lasers with  beatnote $\omega_1-\omega_2=\omega_{\rm t}+\Delta$, such that $q_{\rm d}=({\bf k}_1-{\bf k}_2)\cdot{\bf e}_{\rm z}$ depends on the laser wavevectors projected along the chain, and   $\Omega_{\rm d}/2\pi\sim$1-10kHz. Note that   the crossed-beam ac-Stark shifts must coincide for each  atomic level forming the spin. For cryogenically-cooled SCs~\cite{QS_cirquit_QED}, the bosons are the photonic excitations of  superconducting  resonators of frequency $\omega_{\rm r}/2\pi\sim$1-10GHz, which are  capacitively-coupled yielding  nearest-neighbor tunnelings  $J_{\rm b}a_i^{\dagger}a_{i+1}$, where $J_{\rm b}/2\pi\sim$1-100MHz. A  microwave drive, detuned from the resonator frequency $\omega_{\rm d}=\omega_{\rm r}+\Delta$, is injected in each resonator, and its amplitude/phase is individually controlled by quadrature mixers providing $\Omega_{\rm d}/2\pi\sim$1-100 GHz and a site-dependent phase $\varphi_i=q_{\rm d}di$.

We now introduce the spin-1/2 degrees of freedom. For TIs, two levels are selected from either the hyperfine groundstate  or  a dipole-forbidden optical transition~\cite{QS_trapped_ions}. Using lasers,  a Jaynes-Cummings coupling $g\sigma_i^+a_i\ee^{-\ii\delta t}$ can be introduced, where $g/2\pi\sim$1-50kHz, and $\delta/2\pi\sim$0-0.1MHz is the red-sideband detuning. For SCs, two states with different values of charge/flux variables are separated from the rest by exploiting the Josephson effect~\cite{devoret_review}. These spins can be coupled to the resonator photons via the same Jaynes-Cummings terms, albeit reaching $g/2\pi\sim$1-100MHz and $\delta/2\pi\sim$0.1-1GHz. In the dispersive regime $g\ll\delta$, and setting $J_{\rm b}\ll\omega_{\rm t},\omega_{\rm r}$, these Jaynes-Cummings couplings are highly off-resonant, and lead to  spin-spin $J_{\rm s}= g^2J_{\rm b}/\delta^2$ and spin-boson $U=- g^2/\delta$ interactions to second order, with the peculiarity that TIs yield a dipolar decay  $J_{\rm s}/|i-j|^3$~\cite{jc_lattice_disorder}, while nearest-neighbor couplings are the leading contribution in SCs. 

Let us note that the equivalence of the TI  or SC Hamiltonian  to~\eqref{eq:H0}-\eqref{eq:V} occurs in a rotating frame, where $\omega_{\rm b}=\Delta$  is  the  detuning of the driving, and not the trap or resonator frequencies  $\omega_{\rm t},\omega_{\rm r}$. In order to avoid spurious cross terms  of the driving with the  Jaynes-Cummings terms, we should also impose $g\ll |\delta-\Delta|$. We remark that these realistic experimental parameters allow for a wide tunability of the LL parameters. 

Once the driven spin-boson Hamiltonian is implemented, and the groundstate adiabatically prepared for a certain $K$ of Fig.~\ref{fig_K_parameter}, one must probe the two-point correlators in Fig.~\ref{fig_correlations}. TIs excel at measuring any spin correlator through  site-resolved spin-dependent fluorescence~\cite{QS_trapped_ions}, whereas SCs are better suited to measure the photonic correlations by collecting the output from the cables used to drive the resonators. We thus conclude that either TIs or SCs are promising candidates to realize this spin-boson LL-liquid.

{\it Conclusions.--} We have presented a theoretical study for a new class of spin-boson LLs based on a variational bosonization approach benchmarked by MPS numerics,  and proposed its implementation with TI and SC technologies. This model offers a flexible platform to test qualitatively LL-theory, and displays certain regimes where the LL quantum simulator can beat  MPS numerics on any classical computer.
\ifcheckpagelimits
\else

\acknowledgements
 The authors acknowledge support from EU Project PROMISCE, Spanish MINECO Project FIS2012-33022, and CAM regional research consortium QUITEMAD S2009-ESP-1594.

\clearpage

\begin{widetext}

\hypertarget{sm}{\section{Supplemental Material to ``Driven Spin-Boson Luttinger Liquids''}}

\appendix

\begingroup
\hypersetup{linkcolor=black}
\tableofcontents
\endgroup

\section{Variational minimization and effective long-range XXZ model}
In this part of the Supplemental Material, we describe the variational minimization leading to Eqs.~\eqref{long_range_XXZ} and~\eqref{eff_parameters} of the main text. We first express the original spin-boson Hamiltonian~\eqref{eq:H0}-\eqref{eq:V} in terms of the Jordan-Wigner fermions, and the momentum-space bosons, as defined in the main text. This leads to 
\beq
H=-\frac{\omega_{\rm s}}{2}N-J_{\rm s}\sum_i\left(c_i^{\dagger}c_{i+1}^{\phantom{\dagger}}+{\rm H.c.}\right)+\sum_q\omega_qa_q^{\dagger}a_q^{\phantom{\dagger}}-\Omega_{\rm d}\sqrt{N}\big(a_{q_{\rm d}}^{\dagger}+a_{q_{\rm d}}^{\phantom{\dagger}}\big)+\frac{2U}{N}\sum_i\sum_{q,q'}\ee^{-\ii(q-q')di}c_i^{\dagger}c_{i}^{\phantom{\dagger}}a_q^{\dagger}a_{q'}^{\phantom{\dagger}}.
\eeq
In order to minimize over the variational family of groundstates introduced in the main text, and  given by 
\beq
E_{\rm G}={\rm min}_{\{\alpha_{\rm q},\psi_{\rm f}\}}\left\{ \bra{0_{a}}\bra{{\psi}_{\rm f}}U_{\rm P}U_{\rm B}HU_{\rm B}^{\dagger}U_{\rm P}^{\dagger}\ket{0_{a}}\ket{{\psi}_{\rm f}} \right\}, \hspace{2ex}U_{\rm B}=\ee^{-\sum_{q}(\alpha_qa_q^{\dagger}-{\rm H.c.})}, \hspace{1ex} U_{\rm P}=\ee^{\sum_{q,i}(f_{iq}a_q^{\dagger}-{\rm H.c.})c_i^\dagger c_i^{\phantom{\dagger}}},
\eeq
 we first apply the displacement operator $U_{\rm B}a_qU_{\rm B}^{\dagger}=a_q+\alpha_q$, and proceed in the spirit of the Bogoliubov theory of superfluids~\cite{bogoliubov_weakly_interacting}. Accordingly, we linearise the bosonic density in the Fermi-Bose interaction assuming sufficiently weak interactions, and obtain directly the variational parameter $\alpha_q=\sqrt{N}(\Omega_{\rm d}/\omega_q)\delta_{q,q_{\rm d}}$. For this particular choice, the linear terms in the Fermi-Bose Hamiltonian vanish, and we obtain the following variational problem
$
E_{\rm G}={\rm min}_{\{\psi_{\rm f}\}}\{ \bra{0_{a}}\bra{{\psi}_{\rm f}}U_{\rm P}\tilde{H}U_{\rm P}^{\dagger}\ket{0_{a}}\ket{{\psi}_{\rm f}} \}
$, where 
\beq
\tilde{H}=-\left(\frac{\omega_{\rm s}}{2}+\frac{\Omega_{\rm d}^2}{\omega_{q_{\rm d}}}\right)N-J_{\rm s}\sum_i\left(c_i^{\dagger}c_{i+1}^{\phantom{\dagger}}+{\rm H.c.}\right)+2U\left(\frac{\Omega_{\rm d}}{\omega_{q_{\rm d}}}\right)^2 \sum_ic_i^{\dagger}c_{i}^{\phantom{\dagger}}+\sum_q\omega_qa_q^{\dagger}a_q^{\phantom{\dagger}}+\frac{2U\Omega_{\rm d}}{\sqrt{N}\omega_{q_{\rm d}}}\sum_{i,q}\left(\ee^{-\ii(q-q_{\rm d})di}c_i^{\dagger}c_{i}^{\phantom{\dagger}}a_q^{\dagger}+{\rm H.c.}\right).
\eeq
As noted in the main text, the spin-boson interaction reduces to a Holstein-type coupling~\cite{holstein} within this variational formalism, which corresponds to the last term in the above equation. We note that, contrary to the Holstein model, the Fermi-Bose coupling strength here depends on momentum and the bosons are dispersive. We can now identify the  parameters of the polaron unitary by following the premise that the Fermi-Bose coupling must vanish in the transformed picture. This leads to the polaron parameters $f_{iq}=u_0\ee^{-\ii(q-q_{\rm d})di}/\omega_q\sqrt{N}$ with $u_0=2U\Omega_{\rm d}/\omega_{q_{\rm d}}$. Using the polaron transformation rules
\beq
U_{\rm P}a_qU_{\rm P}^{\dagger}=a_q-\sum_if_{iq} c_i^{\dagger}c_{i}^{\phantom{\dagger}},\hspace{2ex} U_{\rm P}c_{i}^{{\dagger}}U_{\rm P}^{\dagger}=c_{i}^{{\dagger}}\Pi_{q} D_{a_q}(f_{iq})
\eeq
where  $D_{a}(\alpha)={\rm exp}\big(\alpha a^{\dagger}-\alpha^* a\big)$ is the bosonic displacement operator, and taking the expectation value over the  bosonic vacuum, one reduces the variational problem to a purely fermionic one $
E_{\rm G}={\rm min}_{\{\psi_{\rm f}\}}\{ \bra{{\psi}_{\rm f}}H_{\rm f}\ket{{\psi}_{\rm f}} \}
$. Here, $H_{\rm f}$ is  an effective long-range XXZ Hamiltonian described in Eq.~\eqref{long_range_XXZ} of the main text, rewritten here for convenience
\begin{equation}
H_{\rm f}=\sum_i\left(-\tilde{J}_{\rm s}c_i^\dagger c_{i+1}^{\phantom{\dagger}}+{\rm H.c.}\right)
+{\omega}_sn_i+\sum_{i}\sum_{j>i}
{V}_{i-j}n_in_j,
\end{equation}
where $n_i=c_i^\dagger c_i^{\phantom{\dagger}}$ is the fermion number operator, and 
we have introduced the following microscopic parameters
\begin{equation}
V_{i-j}=-\sum_q\omega_q\left(f_{iq}f_{jq}^*+{\rm c.c.}\right),\hspace{2ex} \omega_{\rm s}=2U\left(\frac{\Omega_{\rm d}}{\omega_{q_{\rm d}}}\right)^2+\frac{ V_{0}}{2},\hspace{2ex}\tilde{J}_{\rm s}=J_{s}\bra{0_{a}}\Pi_{q,q'} D_{a_q}(f_{iq})D_{a_{q'}}(-f_{i+1q'})\ket{0_{a}}.
\end{equation}

In order to obtain closed expressions for the parameters of this model, let us introduce $\lambda=2J_{\rm b}/(\omega_{\rm b}-U)$ and assume that $\lambda<1$ to guarantee the absence of negative boson frequencies. We can then use the geometric Taylor series, such that 
\beq
V_{i-j}=-\frac{u_0^2}{(\omega_{\rm b}-U)N}\sum_q\sum_{n=0}^{\infty}\lambda^n\ee^{-\ii(q-q_{\rm d})d(i-j)}\cos^n(qd),
\eeq
together with the binomial theorem after introducing the binomial coefficients
\beq \cos^n(qd)=\frac{1}{2^n}\sum_{k=0}^n\ee^{\ii n qd}\left(\begin{array}{c}n \\k\end{array}\right)\ee^{-\ii2kqd},
\label{binomial}
\eeq
and apply the identity $\sum_q\ee^{\ii q x}=N\delta_{x,0}$. This allows us to express the interaction strengths as a single series, which can be exactly summed by considering a number of combinatorial identities. In this way, we find
\beq
V_{i-j}=-\frac{2u_0^2}{(\omega_{\rm b}-U)\sqrt{1-\lambda^2}}\cos\left(q_{\rm d}d|i-j|\right)\ee^{|i-j|\log\left((1-\sqrt{1-\lambda^2})/\lambda\right)},
\eeq
which yields the expression in Eq.~\eqref{eff_parameters} of the main text with the parameters listed in~\cite{explicit_int}. The validity of this derivation relies on the exact agreement with the numerical estimates displayed in Figs.~\ref{fig_parameters_XXZ}{\bf (a)},{\bf (c)}, and {\bf (d)}. 

We now  use the identity for the overlap of coherent states $\ket{\alpha}=D_a(\alpha)\ket{0_a}$, namely $\langle\alpha|\beta\rangle=\ee^{-\half(\alpha^*\beta-\beta^*\alpha)}\ee^{-\half|\alpha-\beta|^2}$. This allows us to express the dressed tunnelling as an exponential renormalization of the bare tunnelling $\tilde{J}_{\rm s}=J_{s}\ee^{-\chi}$ with
\beq
\chi=\sum_q\frac{u_0^2}{N\omega_q^2}\left(1-\ee^{\ii(q-q_{\rm d})d}\right).
\label{chi}
\eeq
This sum can be evaluated following a similar procedure as above. We first use the binomial Taylor series, such that 
\beq
\chi=\frac{u_0^2}{(\omega_{\rm b}-U)^2N}\sum_q\sum_{n=0}^\infty (n+1)\lambda^n\left(1-\ee^{\ii(q-q_{\rm d})d}\right)\cos^n(qd).
\eeq
This expression can be analytically summed by using again Eq.~\eqref{binomial} , together with some combinatorial identities, such that
\beq
\tilde{J}_{\rm s}=J_{\rm s}\ee^{-\frac{u_0^2}{(\omega_{\rm b}-U)^2}\frac{\left(1-\lambda\ee^{-\ii q_{\rm d}d}\right)}{(1-\lambda^2)^{3/2}}},
\label{dressed_tunneling}
\eeq
which coincides with Eq.~\eqref{eff_parameters} of the main text with the parameters listed in~\cite{explicit_tunn}. The validity of this derivation has been confronted to the exact numerics of Fig.~\ref{fig_parameters_XXZ}{\bf (b)}.

\section{Constructive bosonization of the spin-boson lattice model}
In this part of the Supplemental Material, we present the details for the derivation of variational bosonization ansatz of Eqs.~\eqref{bosonization}-\eqref{bosonized_var_gs} in the main text, and the corresponding two-point correlators in Eqs.~\eqref{szsz_LL}-\eqref{ada_LL}.

 We start by setting  ${\omega}_{\rm s}=0$ in Eq.~\eqref{long_range_XXZ}, which is reasonable for sufficiently  weak interactions such that $\tilde{J}_{\rm s}\gg{\omega}_{\rm s}$. Assuming periodic boundary conditions,  the kinetic part of the  fermionic Hamiltonian $H_{\rm f}=K_{\rm f}+V_{\rm f}$ can be written as 
 \beq 
 K_{\rm f}=-|\tilde{J}_{\rm s}|\sum_ic_i^\dagger (c_{i+1}+c_{i-1}),
 \eeq
 where the phase of the dressed tunnelling $\ee^{\ii {\rm arg}(\tilde{J}_{\rm s})}$ has been absorbed in the fermionic operators via a  $U(1)$ gauge transformation.   This  yields a  band structure $\epsilon_{\rm s}(q)=2|\tilde{J}_{\rm s}|\cos(qd)$,
where  groundstate is obtained by filling all  negative-energy levels $-k_{\rm F}\leq q\leq k_{\rm F}$, with $k_{\rm F}=\pi/2d$, and the low-energy excitations correspond to right- and left-moving fermions $\eta\in\{\rm R,L\}$ with momentum close to  $\pm k_{\rm F}$, respectively. In the continuum limit, we let $d\to 0$ and $N\to\infty$ such that the length $L=Nd$ remains constant. The   low-energy properties  are described by  a continuum  field theory    of slowly-varying fields for the left/right-moving fermions
\beq
c_i={\sqrt{\frac{d}{2\pi}}}\Psi(x),\hspace{1ex}c_{i\pm1}={\sqrt{\frac{d}{2\pi}}}\Psi(x\pm d),\hspace{3ex}
\Psi(x)=\ee^{\ii k_{\rm F}x}\tilde{\Psi}_{\rm R}(x)+\ee^{-\ii k_{\rm F}x}\tilde{\Psi}_{\rm L}(x).
\label{continuum}
\eeq 
After Taylor expanding the slowly-varying fields $\tilde{\Psi}_{\eta}(x\pm d)\approx\tilde{\Psi}_{\eta}(x)\pm d\partial_x\tilde{\Psi}_{\eta}(x)+\half d^2\partial^2_x\tilde{\Psi}_{\eta}(x)\pm \dots,$   the kinetic energy corresponds to a $1+1$ Dirac quantum field theory for massless fermions
\beq
K_{\rm f}= \frac{v_{\rm F}}{2\pi}\int_{-\frac{L}{2}}^{\frac{L}{2}}{\rm d}x\left( \tilde{\Psi}_{\rm L}^{\dagger}(x)\ii\partial_x\tilde{\Psi}_{\rm L}(x)-\tilde{\Psi}_{\rm R}^{\dagger}(x)\ii\partial_x\tilde{\Psi}_{\rm R}(x)\right),\label{dirac}
\eeq
 where   rapidly-oscillating terms $\ee^{\pm \ii 2k_{\rm F}x}=(-1)^{x/d}$ average out under the integral, higher-order derivatives can be neglected for $d\to 0$, and we have introduced the Fermi velocity  that plays the role of the speed of light in the continuum limit $2|\tilde{J}_{\rm s}|d\to v_{\rm F}$. 
 
 The first step of the  constructive  bosonization~\cite{const_bosonization} is to extend this description, which is expected to be valid around the Fermi points, to all possible momenta $k=\frac{2\pi}{L}n_{k}$ where $n_{k}\in\mathbb{Z}$ (i.e. the fermionic spectrum is linearized for all momenta), such that 
 \beq
 \tilde{\Psi}_{\rm R}(x)=\sqrt{\frac{2\pi}{L}}\sum_{k=-\infty}^{\infty}\ee^{\ii kx}c_{k,{\rm R}},\hspace{3ex} \tilde{\Psi}_{\rm L}(x)=\sqrt{\frac{2\pi}{L}}\sum_{k=-\infty}^{\infty}\ee^{-\ii kx}c_{k,{\rm L}},\hspace{3ex} K_{\rm f}=\sum_{k,\eta}v_{\rm F}k:c_{k,\eta}^\dagger c_{k,\eta}:
 \label{fermion_fields}
 \eeq
where we have introduced the normal ordering $:(\hspace{1 ex}):$ with respect to the groundstate of the free theory. In this way,  we can focus on  certain low-energy excitations above the Dirac sea of filled negative-energy states by defining  operators associated to the particle-hole excitations of momentum $q=\frac{2\pi}{L}n_{q}>0$ for the left/right fermionic branches, namely
 \beq 
 b_{q\eta}=\frac{-\ii}{\sqrt{n_q}}\sum_{k=-\infty}^{\infty}c_{k-q,\eta}^\dagger c_{k,\eta}=( b_{q\eta}^{\dagger})^{\dagger},
 \eeq
 which turn out to be bosonic.
 From the commutation rules of these  operators with the Dirac Hamiltonian~\eqref{dirac}, it follows that 
$
 K_{\rm f}=\sum_{q}v_{\rm F}q (b_{q{\rm R}}^\dagger b_{q{\rm R}}+b_{q{\rm L}}^\dagger b_{q{\rm L}})
$
 in the thermodynamic limit $L\to \infty$. Therefore, the low-energy excitations of the free fermionic  model are described by  two  bosonic branches for the particle-hole excitations around each Fermi point (i.e. {\it bosonization}). 
 
 It is customary to express the Hamiltonian in terms of bosonic fields
 \beq
  {\phi}_{\rm R}(x)=\varphi_{\rm R}(x)+{\rm H.c.}, \hspace{1ex}\varphi_{\rm R}(x)=\sum_{n_q>0}\frac{-1}{\sqrt{n_q}}\ee^{\ii qx-\frac{a q}{2}} b_{q{\rm R}},\hspace{3ex} \ {\phi}_{\rm L}(x)=\varphi_{\rm L}(x)+{\rm H.c.},  \hspace{1ex} \varphi_{\rm L}(x)=\sum_{n_q>0}\frac{-1}{\sqrt{n_q}}\ee^{-\ii qx -\frac{a q}{2}} b_{q{\rm L}},
  \label{bosonic_fields}
   \eeq
 where $a>0$ is a regularization constant that cuts off large momenta to ensure convergence. Since we are interested in low-energy properties, this cutoff does not change the physics, and we can set it to zero at the end of the calculations. The kinetic energy, and thus the 1+1 Dirac Hamiltonian,  can then be expressed as a free bosonic field theory
 \beq
 K_{\rm f}=\int_{-\infty}^{\infty}\frac{{\rm d}x}{2\pi}\frac{v_{\rm F}}{2}\bigg( :(\partial_x{\phi}_{{\rm R}}(x))^2:+:(\partial_x{\phi}_{\rm L}(x))^2:\bigg)=\int_{-\infty}^{\infty}\frac{{\rm d}x}{2\pi}\frac{v_{\rm F}}{2}\bigg( :(\partial_x{\Phi}(x))^2:+:(\partial_x{\Theta}(x))^2:\bigg),
 \label{boson_kinetic}
 \eeq
 where we have also introduced the so-called dual fields $\Theta(x)=(\phi_{\rm R}(x)+\phi_{\rm L}(x))/\sqrt{2}$, and $\Phi(x)=(\phi_{\rm R}(x)-\phi_{\rm L}(x))/\sqrt{2}$. 
 
 The equivalence of the fermionic~\eqref{dirac} and bosonic~\eqref{boson_kinetic}  Hamiltonians suggests the existence of a direct mapping between fermionic and bosonic fields. To obtain the correct fermionic  anticommutation relations, one has to introduce the so-called Klein factors $F_{\eta}$ fulfilling $[F_{\eta},\phi_{\eta'}]=0$,  $\{F^\dagger_{\eta},F_{\eta'}\}=2\delta_{\eta,\eta'}$, and $\{F_{\eta},F_{\eta'}\}=0$ if $\eta\neq\eta'$.
 The  bosonization identity~\cite{const_bosonization} then relates the fermionic and bosonic fields in the thermodynamic limit as follows 
 \beq
  \tilde{\Psi}_{\rm R}(x)=\frac{F_{\rm R}}{\sqrt{a}}\ee^{-\ii\phi_{\rm R}(x)}=\sqrt{\frac{2\pi}{L}}F_{\rm R}\ee^{-\ii\varphi^\dagger_{\rm R}(x)}\ee^{-\ii\varphi_{\rm R}(x)},\hspace{3ex}   \tilde{\Psi}_{\rm L}(x)=\frac{F_{\rm L}}{\sqrt{a}}\ee^{-\ii\phi_{\rm L}(x)}=\sqrt{\frac{2\pi}{L}}F_{\rm L}\ee^{-\ii\varphi^\dagger_{\rm L}(x)}\ee^{-\ii\varphi_{\rm L}(x)}.
  \label{bos_id}
  \eeq
Equipped with this operator identity, we can  bosonize the  interaction~\eqref{long_range_XXZ} which,  in terms of the fermion fields~\eqref{continuum}, reads
 \beq
 V_{\rm f}=\sum_{\ell\geq 1}\frac{\tilde{V}_{\ell}}{2\pi}\int_{-\frac{L}{2}}^{\frac{L}{2}}\frac{{\rm d}x}{2\pi}\Psi^\dagger(x)\Psi(x)\Psi^{\dagger}(x+\ell d)\Psi(x+\ell d),
 \eeq
 where we must define $V_{\ell}d\to\tilde{V}_{\ell}$ in the continuum limit in analogy to the Fermi velocity below Eq.~\eqref{dirac}. The bosonization of these interactions is more intricate, as one must  avoid possible divergences by  normal ordering. Besides, the long-range tail of the interaction allows for $\ell\to\infty$ in the continuum limit, such that special care must be taken in the truncation of the  Taylor series of the  fermionic fields for $d\to 0$. This will impose restrictions on the interaction range tractable by bosonization.

Assuming that the interactions are small enough $\tilde{V}_{\ell}\ll v_{\rm F}$, such that the slowly-varying fields~\eqref{continuum} are only slightly perturbed, we  identify the  interaction terms $ V_{\rm f}= V_{\rm f}^0+ V_{\rm f}^{2k_{\rm F}}+ V_{\rm f}^{4k_{\rm F}}$, where we have again neglected  rapidly-oscillating contributions
 \beq
 \begin{split}
  V_{\rm f}^0&=\sum_{\ell\geq 1}\frac{\tilde{V}_{\ell}}{2\pi}\int_{-\frac{L}{2}}^{\frac{L}{2}}\frac{{\rm d}x}{2\pi}\sum_{\eta,\eta'} :\tilde{\Psi}_\eta^\dagger(x) \tilde{\Psi}_{\eta}(x) \tilde{\Psi}_{\eta'}^{\dagger}(x+\ell d) \tilde{\Psi}_{\eta'}(x+\ell d):,\\
     V_{\rm f}^{2k_{\rm F}}&=\sum_{\ell\geq 1}(-1)^\ell\frac{\tilde{V}_{\ell}}{2\pi}\int_{-\frac{L}{2}}^{\frac{L}{2}}\frac{{\rm d}x}{2\pi}\sum_{\eta\neq\eta'} :\tilde{\Psi}_\eta^\dagger(x) \tilde{\Psi}_{\eta'}(x) \tilde{\Psi}_{\eta'}^{\dagger}(x+\ell d) \tilde{\Psi}_{\eta}(x+\ell d):,\\
         V_{\rm f}^{4k_{\rm F}}&=\sum_{\ell\geq 1}(-1)^\ell\frac{\tilde{V}_{\ell}}{2\pi}\int_{-\frac{L}{2}}^{\frac{L}{2}}\frac{{\rm d}x}{2\pi}\sum_{\eta\neq\eta'} :\tilde{\Psi}_\eta^\dagger(x) \tilde{\Psi}_{\eta'}(x) \tilde{\Psi}_{\eta}^{\dagger}(x+\ell d) \tilde{\Psi}_{\eta'}(x+\ell d):.
  \end{split}
 \eeq
 
 These terms are expressed in terms of the bosonic fields by means of the  identity~\eqref{bos_id}. To eliminate  possible divergences in the first term, we use the so-called point-splitting regularisation $:\tilde{\Psi}_{\eta}^{\dagger}(x)\tilde{\Psi}_{\eta}(x):=\tilde{\Psi}_{\eta}^{\dagger}(x+\epsilon)\tilde{\Psi}_{\eta}(x)-\bra{0}\tilde{\Psi}_{\eta}^{\dagger}(x+\epsilon)\tilde{\Psi}_{\eta}(x)\ket{0}$, where $\ket{0}$ is a reference state without particle-hole excitations, and $\epsilon\to 0$ such that we can Taylor expand the field operators. Accordingly, we find $
:\tilde{\Psi}_\eta^\dagger(x) \tilde{\Psi}_{\eta}(x):=(-\delta_{\eta,{\rm R}}+\delta_{\eta,{\rm L}})\partial_x\phi_{\eta}$, which allows us to bosonize the first   $V_{\rm f}^0$ interaction term
 \beq
\sum_{\eta,\eta'}:\tilde{\Psi}_\eta^\dagger(x) \tilde{\Psi}_{\eta}(x) \tilde{\Psi}_{\eta'}^{\dagger}(x+\ell d) \tilde{\Psi}_{\eta'}(x+\ell d):=:(\partial_x\phi_{\rm R}(x)-\partial_x\phi_{\rm L}(x))(\partial_x\phi_{\rm R}(x+\ell d)-\partial_x\phi_{\rm L}(x+\ell d)):.
 \eeq
 We can now Taylor expand  the bosonic fields ${\phi}_{\eta}(x+ \ell d)\approx{\phi}_{\eta}(x)+ \ell d\partial_x{\phi}_{\eta}(x)+\dots$ and, provided that the interactions decay fast enough, namely
 \beq
\lim_{|i-j|\to \infty}\tilde{V}_{i-j} \cdot |i-j| \to C\equiv{\rm constant},
\label{decay}
 \eeq
 the corresponding fermionic quartic term can be expressed  as a quadratic bosonic one  in the $d\to 0$ limit
  \beq
\sum_{\eta,\eta'}:\tilde{\Psi}_\eta^\dagger(x) \tilde{\Psi}_{\eta}(x) \tilde{\Psi}_{\eta'}^{\dagger}(x+\ell d) \tilde{\Psi}_{\eta'}(x+\ell d):=:(\partial_x\phi_{\rm R}(x)-\partial_x\phi_{\rm L}(x))^2:.
 \eeq
  This means that  only interactions that decay faster than, or equal  to, the Coulomb interaction~\eqref{decay} can be treated via the constructive bosonization. This is crucial to neglect higher-order derivatives, which scale with $C\ell^{n-1}d^n\to 0$ in the continuum limit provided that Eq.~\eqref{decay} is fulfilled, even for $\ell\to\infty$ in the thermodynamic limit. It is also clear that the bosonization predictions will be more accurate the faster the decay is, since the contributions of the higher-order derivatives will be less and less  important.
 
 We can proceed similarly with the second term $     V_{\rm f}^{2k_{\rm F}}$. Using the Baker-Campbell-Hausdorff formula with   $[\varphi^\dagger_\eta(x),\varphi_{\eta'}(x')]=-\delta_{\eta,\eta'}(\delta_{\eta,{\rm R}}\log\big(\frac{2\pi}{L}(a+ \ii (x-x')\big)+\delta_{\eta,{\rm L}}\log\big(\frac{2\pi}{L}(a- \ii (x-x')\big))$, which can be checked from Eq.~\eqref{bosonic_fields}, we can write the second interaction term  in the $a\to 0$, and $d\to 0$, limit as follows
 \beq
     \sum_{\eta\neq\eta'} :\tilde{\Psi}_\eta^\dagger(x) \tilde{\Psi}_{\eta'}(x) \tilde{\Psi}_{\eta'}^{\dagger}(x+\ell d) \tilde{\Psi}_{\eta}(x+\ell d):=\frac{1}{(\ell d)^2}:\ee^{-\ii(\ell d)\partial_x(\varphi_{\rm R}^\dagger(x)-\varphi_{\rm L}^\dagger(x))}\ee^{-\ii(\ell d)\partial_x(\varphi_{\rm R}(x)-\varphi_{\rm L}(x))}:+{\rm H.c.}.
 \eeq
 After Taylor expansion, and making use of the constraint on the interaction decay~\eqref{decay}, we find
 \beq
  \sum_{\eta\neq\eta'} :\tilde{\Psi}_\eta^\dagger(x) \tilde{\Psi}_{\eta'}(x) \tilde{\Psi}_{\eta'}^{\dagger}(x+\ell d) \tilde{\Psi}_{\eta}(x+\ell d):=-:(\partial_x\phi_{\rm R}(x)-\partial_x\phi_{\rm L}(x))^2:.
 \eeq
 The last term $        V_{\rm f}^{4k_{\rm F}}$, which corresponds to the so-called Umklapp scattering, is also bosonized by using the Baker-Campbell-Hausdorff formula and a Taylor expansion, such that we find in the limit $d\to 0$, we obtain
 \beq
 \sum_{\eta\neq\eta'} :\tilde{\Psi}_\eta^\dagger(x) \tilde{\Psi}_{\eta'}(x) \tilde{\Psi}_{\eta}^{\dagger}(x+\ell d) \tilde{\Psi}_{\eta'}(x+\ell d):=\frac{1}{a^2}:(F_{\rm R}^\dagger F_{\rm L})^2\ee^{\ii2\phi_{\rm R}(x)}\ee^{-\ii2\phi_{\rm L}(x)}:+{\rm H.c.}\equiv\frac{2}{a^2}:\cos(2\phi_{\rm R}(x)-2\phi_{\rm L}(x)):,
 \eeq
 where the definition of the cosine operator  in the last term must incorporate the combination of Klein factors.
 
Using all these expressions, it is possible to rewrite the  fermionic Hamiltonian as a {\it sine-Gordon quantum  field theory} composed of the Hamiltonian of a Luttinger liquid $H_{\rm LL}$, and a non-linearity due to the Umklapp scattering $H_{\rm U}$,
\beq
H_{\rm f}=H_{\rm LL}+H_{\rm U},\hspace{2ex}H_{\rm LL}=\int_{-\infty}^{\infty}\frac{{\rm d}x}{2\pi}\frac{u}{2}\bigg( \frac{1}{K}:(\partial_x{\Phi}(x))^2:+K:(\partial_x{\Theta}(x))^2:\bigg), \hspace{3ex} H_{\rm U}=\int_{-\infty}^{\infty}\frac{{\rm d}x}{2\pi}g_{\rm U}:\cos(2\sqrt{2}\Phi(x)):,
\label{sine_gordon}
\eeq 
where $u$ is a renormalized Fermi velocity,  $K$ is the so-called Luttinger parameter, and $g_{\rm U}$ the Umklapp interacting strength. For  the small interactions $\tilde{V}_{\ell}\ll v_{\rm F}$ considered here, these parameters are obtained through the above expressions
\beq
u=v_{\rm F}\left(1+\sum_{\ell\geq 1}\frac{\tilde{V}_{\ell}(1-(-1)^\ell)}{\pi v_{\rm F}}\right)^{\half},\hspace{3ex} K=\left(1+\sum_{\ell\geq 1}\frac{\tilde{V}_{\ell}(1-(-1)^\ell)}{\pi v_{\rm F}}\right)^{-\half},\hspace{3ex} g_{\rm U}=\sum_{\ell\geq 1}\frac{\tilde{V}_{\ell}(-1)^\ell}{\pi a^2},
\label{LL_parameters}
\eeq
which  coincide with  a phenomenological bosonization~\cite{capponi}  up to a different choice of fermionic algebra and dual fields. Using the exact expression of the mediated interactions in Eq.~\eqref{eff_parameters}, we can find the analytical expressions for the Luttinger parameters by using the geometric Taylor series and some trigonometric identities, such that 
\beq
\begin{split}
&\sum_{\ell=1}^{\infty}\tilde{V}_{\ell}=\frac{\tilde{V}_0}{2}\sum_{\ell=1}^{\infty}\left(\ee^{(\ii q_{\rm d}-\xi_0^{-1})d}\right)^\ell+\frac{\tilde{V}_0}{2}\sum_{\ell=1}^{\infty}\left(\ee^{(-\ii q_{\rm d}-\xi_0^{-1})d}\right)^\ell=\frac{\tilde{V}_0}{2}\frac{\cos(q_{\rm d}d)-\ee^{-d/\xi_0}}{{\rm cosh}(d/\xi_0)-\cos(q_{\rm d}d)},\\
&\sum_{\ell=1}^{\infty}(-1)^\ell\tilde{V}_{\ell}=\frac{\tilde{V}_0}{2}\sum_{\ell=1}^{\infty}\left(-\ee^{(\ii q_{\rm d}-\xi_0^{-1})d}\right)^\ell+\frac{\tilde{V}_0}{2}\sum_{\ell=1}^{\infty}\left(-\ee^{(-\ii q_{\rm d}-\xi_0^{-1})d}\right)^\ell=\frac{\tilde{V}_0}{2}\frac{-\cos(q_{\rm d}d)-\ee^{-d/\xi_0}}{{\rm cosh}(d/\xi_0)+\cos(q_{\rm d}d)}.\\
\end{split}
\eeq
These expressions yield the Luttinger parameters in Eqs.~\eqref{u_parameter} and~\eqref{K_parameter} of the main text. 

 For small-enough interactions, the sine-Gordon non-linearity is irrelevant~\cite{giamarchi_book}, and the bosonized groundstate is solely determined by the Luttinger-liquid Hamiltonian $H_{\rm f}\approx H_{\rm LL}$ in Eq.~\eqref{sine_gordon}. To obtain the corresponding  groundstate, let us express the dual fields in terms of new bosonic fields $\Theta(x)=(\phi_{+}(x)+\phi_{-}(x))\sqrt{1/2K}$, and $\Phi(x)=(\phi_{+}(x)-\phi_{-}(x))\sqrt{K/2}$, where
 \beq
  {\phi}_{\rm +}(x)=\sum_{n_q>0}\frac{-1}{\sqrt{n_q}}\ee^{-\ii qx-\frac{a q}{2}} d_{q{\rm +}}+{\rm H.c.},\hspace{3ex}   {\phi}_{\rm -}(x)=\sum_{n_q>0}\frac{-1}{\sqrt{n_q}}\ee^{+\ii qx-\frac{a q}{2}} d_{q{\rm -}}+{\rm H.c.},
  \label{squeezed_bosonic_fields}
   \eeq
are defined in terms of two species of bosonic creation-annihilation operators $d_{q{\pm}}^{\dagger},d_{q{\pm}}$. By direct substitution, one finds that in analogy to the free theory,
the spectrum of the full interacting theory can again be described by two linear bosonic branches 
\beq
H_{\rm f}=\sum_{q}uq (d_{q{\rm +}}^\dagger d_{q{\rm +}}+d_{q{\rm -}}^\dagger d_{q{\rm -}}), 
\eeq
with a renormalized Fermi velocity $v_{\rm F}\to u$~\eqref{LL_parameters}, which proves Eq.~\eqref{bosonization} in the main text. Therefore, the  fermionic part of the variational groundstate  is the vacuum of the new bosonic modes $\ket{\psi_{\rm f}}=\ket{0_d}$, and directly leads to the complete  variational bosonization ansatz of  Eq.~\eqref{bosonized_var_gs}, namely  $\ket{\Psi_{\rm G}(\{\alpha_q,{\psi}_{\rm f}\})}=U_{\rm B}^\dagger U_{\rm P}^\dagger\ket{0_a}\otimes\ket{0_d}.$

Once the variational bosonization ansatz is known, we can embark upon the calculation of the two-point correlators. We start from the simplest one, the diagonal spin-spin correlators $
 C_{\sigma^z\sigma^z}(\ell)=\sum_i(\langle \sigma^z_{\ell+i}\sigma^z_i\rangle-\langle \sigma^z_{\ell+i}\rangle\langle \sigma^z_{i}\rangle)/N$. Making use of the Jordan-Wigner mapping, one finds  
 \beq
  C_{\sigma^z\sigma^z}(\ell)=4\bra{0_a,0_d}U_{\rm P}U_{\rm B}\left(n_{\ell+1}-\half\right)\left(n_{1}-\half\right)U_{\rm B}^\dagger U_{\rm P}^\dagger\ket{0_a,0_d},
  \eeq
  and since both unitaries commute with the fermion  operators, it follows that 
$
  C_{\sigma^z\sigma^z}(x)=\frac{d^2}{\pi^2}\bra{0_d}:\Psi^{\dagger}(x)\Psi^{\phantom{\dagger}}(x)::\Psi^{\dagger}(0)\Psi^{\phantom{\dagger}}(0):\ket{0_d},
$   where we have already taken the continuum limit in Eq.~\eqref{continuum}. This expectation value coincides exactly with that of the bosonized XXZ model~\cite{luther_peschel}, which has become a textbook example~\cite{giamarchi_book}. Using the above bosonization relations, and the Klein-factor algebra, one finds $
  C_{\sigma^z\sigma^z}(x)=\frac{2d^2}{\pi^2}\bra{0_d}\partial_x\Phi(x)\partial_x\Phi(0)\ket{0_d}+\frac{d^2}{\pi^2a^2}(\ee^{-\ii2k_{\rm F}x}\bra{0_d}\ee^{\ii\sqrt{2}\Phi(x)}\ee^{-\ii\sqrt{2}\Phi(0)}\ket{0_d}+{\rm c.c.})$. The first term can be evaluated easily after expressing the dual field in terms of the new bosonic creation-annihilation operators~\eqref{squeezed_bosonic_fields}, while the second term requires the additional identity for Gaussian states $\bra{0_d}\ee^{\ii\sqrt{2}\Phi(x)}\ee^{-\ii\sqrt{2}\Phi(0)}\ket{0_d}=\ee^{2\bra{0_d}\Phi(x)\Phi(0)\ket{0_d}}\ee^{-\bra{0_d}\Phi^2(x)\ket{0_d}}\ee^{-\bra{0_d}\Phi^2(0)\ket{0_d}}$. Altogether, one obtains in the thermodynamic limit $L\to \infty$
 \begin{equation}
 C_{\sigma^z\sigma^z}(x)=-\frac{2d^2}{\pi^2}\left(\frac{K}{x^2}-\ee^{-\ii2k_{\rm F}x}\left(\frac{d}{x}\right)^{2K}\right),
\end{equation}
which corresponds to Eq.~\eqref{szsz_LL} after setting $x=\ell d$, and $2k_{\rm F}d=\pi$.

For the off-diagonal spin correlators 
 $C_{\sigma^+\sigma^-}(\ell)=\sum_i(\langle \sigma^+_{\ell+i}\sigma^-_i\rangle-\langle \sigma^+_{\ell+i}\rangle\langle \sigma^-_{i}\rangle)/N$, we proceed analogously to find 
  \beq
  C_{\sigma^+\sigma^-}(\ell)=\Pi_{q,q'}\bra{0_a}D_{a_q}(f_{1+\ell,q})D_{a_{q'}}(-f_{1,q'})\ket{0_a}\bra{0_d}\sigma_{1+\ell}^+\sigma_1^-\ket{0_d}.
  \eeq
  The bosonic contribution is due to the polaron cloud dressing the spins, and can be calculated in analogy to the renormalization of the tunnelling in Eq.~\eqref{chi} by using the coherent-state algebra. The spin contribution reduces once more to that of the bare XXZ model, which requires additional care due to the Jordan-Wigner string when expressed in terms of fermions~\cite{giamarchi_book}. Anyhow, it can be expressed again as a collection of expectation values of exponentials of the dual fields in the continuum limit, which can be calculated exactly using the same procedure as above. Altogether, this leads to
  \beq
  C_{\sigma^+\sigma^-}(x)=\ee^{-\tilde{\chi}(x)}\frac{1}{2\pi}\left(\frac{d}{x}\right)^{1/2K}\left(1+2\ee^{-\ii 2k_{\rm F}x}\left(\frac{d}{x}\right)^{2K}\right), 
  \eeq
  where $\tilde{\chi}(x)=\sum_q\frac{u_0^2}{N\omega_q^2}\left(1-\ee^{\ii(q-q_{\rm d})x}\right).$ This function can be evaluated by using once more the binomial series together with combinatorial relations, and after setting $x=\ell d$, and $2k_{\rm F}d=\pi$, we find precisely the expression in Eq.~\eqref{spsm_LL} of the main text.

Moving onto the bosons, let us address the off-diagonal  correlators  $C_{a^\dagger a}(\ell)=\sum_i(\langle a_{\ell+i}^\dagger a_i^{\phantom{\dagger}}\rangle-\langle a_{\ell+i}^\dagger \rangle\langle a_i^{\phantom{\dagger}}\rangle)/N$. Using the variational bosonization ansatz, we find that such correlators depend on the spin structure factors ${\rm S}_{\sigma^z\sigma^z}(q)$ as follows
\beq
C_{a^\dagger a}(\ell)=\frac{u_0^2}{4N}\sum_q\frac{1}{\omega_q^2}{\rm S}_{\sigma^z\sigma^z}(q-q_{\rm d})\ee^{\ii qd\ell},\hspace{2ex}{\rm S}_{\sigma^z\sigma^z}(q)=\sum_{\ell'}\ee^{-\ii qd\ell'}C_{\sigma^z\sigma^z}(\ell').
\eeq
To evaluate this expression, we define the  inverse Fourier series $1/\Omega^2_j=\sum_q\ee^{\ii qdj}/\omega_q^2\sqrt{N}$, which can be evaluated again making use of the binomial series and binomial theorem
\beq
\frac{1}{\Omega^2_j}=\frac{1+j\sqrt{1-\lambda^2}}{(1-\lambda^2)^{\frac{3}{2}}}\ee^{-\frac{dj}{\xi_0}}.
\eeq
Using  this expression and the identity  $\sum_q\ee^{\ii q x}=N\delta_{x,0}$, one can express the bosonic correlators as
\beq
C_{a^\dagger a}(\ell)=\frac{u_0^2}{4}\sum_{\ell'}\frac{1}{\Omega_{\ell-\ell'}^2}C_{\sigma^z\sigma^z}(\ell')\ee^{\ii q_{\rm d}\ell'},
\eeq
which leads directly to Eq.~\eqref{ada_LL} upon substitution of the diagonal spin-spin correlators in Eq.~\eqref{szsz_LL}  already derived in this Supplemental Material.
\end{widetext}

\begin{references}

\bibitem{QS_ultracold_atoms}
{\it See} I. Bloch, J. Dalibard and S. Nascimb\'ene, \href{http://www.nature.com/nphys/insight/quantum-simulation/index.html}{Nat. Phys. {\bf 8,} 267 (2012)}, {\it and references therein.} 

\bibitem{QS_trapped_ions}
{\it See} R. Blatt and C. F. Roos, \href{http://www.nature.com/nphys/insight/quantum-simulation/index.html}{Nat. Phys. {\bf 8,} 277 (2012)}, {\it and references therein.} 

\bibitem{QS_cirquit_QED}
{\it See} A.A. Houck, H.E. T\"ureci, and J. Koch, \href{http://www.nature.com/nphys/insight/quantum-simulation/index.html}{Nat. Phys. {\bf 8,} 264 (2012)},  {\it and references therein.} 

\bibitem{feynman}
R. P. Feynman, \href{http://link.springer.com/article/10.1007%2FBF02650179}{Int. J. Theo. Phys. {\bf 21,} 467 (1982).}

\bibitem{QS}
J. I Cirac and P. Zoller, \href{http://www.nature.com/nphys/insight/quantum-simulation/index.html}{Nat. Phys. {\bf 8,} 264 (2012).}

 \bibitem{wen_book}
X.-G. Wen, {\it Quantum Field Theory of Many-body Systems} (Oxford University Press, Oxford, 2004).
 
\bibitem{giamarchi_book}
T. Giamarchi, {\it Quantum Physics in One Dimension} (Oxford University Press, Oxford, 2004).

\bibitem{phenom_bosonization}
F. D. M. Haldane, \href{http://journals.aps.org/prl/abstract/10.1103/PhysRevLett.47.1840}{Phys. Rev. Lett. {\bf 47,} 1840 (1981) }.

\bibitem{spinon_spectrum_magnets}
D.A. Tennant, R.A. Cowley, S.E. Nagler, and A.M. Tsvelik, \href{http://journals.aps.org/prb/abstract/10.1103/PhysRevB.52.13368}{Phys. Rev. B {\bf 52,} 13368 (1995)}; B. Lake, D.A. Tennant, C.D. Frost, and S.E. Nagler, \href{http://www.nature.com/nmat/journal/v4/n4/abs/nmat1327.html}{Nat. Materials {\bf 4,} 329 (2005)}



\bibitem{LL_organic_salts}
A. Schwartz, M. Dressel, G. Gr\"{u}ner, V. Vescoli, L. Degiorgi, and T. Giamarchi, \href{http://journals.aps.org/prb/abstract/10.1103/PhysRevB.58.1261}{Phys. Rev. B {\bf 58,} 1261 (1998).}

\bibitem{LL_carbon_nanotubes}
M. Bockrath, D.H. Cobden, J. Lu, A.G. Rinzler, R.E. Smalley, L. Balents, and P. L. McEuen, \href{http://www.nature.com/nature/journal/v397/n6720/abs/397598a0.html}{Nature {\bf 397,} 598  (1999)}; Z. Yao, H.W.Ch. Postma, L. Balents, and C. Dekker, \href{http://www.nature.com/nature/journal/v402/n6759/full/402273a0.html}{Nature {\bf 402,} 273 (1999)}.

\bibitem{LL_semiconducting_quantum_wires}
O. M. Auslaender, A. Yacoby, R. de Picciotto, K. W. Baldwin, L. N. Pfeiffer, and K. W. West, \href{http://journals.aps.org/prl/abstract/10.1103/PhysRevLett.84.1764}{Phys. Rev. Lett. {\bf 84,} 1764 (2000)}; O. M. Auslaender, A. Yacoby, R. de Picciotto, K.W. Baldwin, L.N. Pfeiffer, K.W. West, \href{http://www.sciencemag.org/content/295/5556/825}{Science {\bf 295,} 825 (2002).}

\bibitem{ll_xxz_magnetic_ladders}
M. Klanjsek, H. Mayaffre, C. Berthier, M. Horvatic, B. Chiari, O. Piovesana, P. Bouillot, C. Kollath, E. Orignac, R. Citro, and T. Giamarchi, \href{http://journals.aps.org/prl/abstract/10.1103/PhysRevLett.101.137207}{Phys. Rev. Lett. {\bf 101,} 137207 (2008).}

\bibitem{giamarchi_exp_review}
T. Giamarchi, \href{http://www.worldscientific.com/doi/abs/10.1142/S0217979212440043?queryID=%24%7BresultBean.queryID%7D}{Int. J. Mod. Phys. B {\bf 26,} 1244004 (2012).} 


\bibitem{cold_bosons_sine_Gordon_qpt_LL}
S. Hofferberth, I. Lesanovsky, T. Schumm, A. Imambekov, V. Gritsev, E. Demler, and J. Schmiedmayer,  \href{http://www.nature.com/nphys/journal/v4/n6/full/nphys941.html}{Nat. Phys. {\bf 4,} 489 (2008)}; E. Haller,	 R. Hart,	 M.J. Mark,	 J.G. Danzl,	 L. Reichs\"{o}llner,	 M. Gustavsson,	 M. Dalmonte,	 G. Pupillo, and H.-C. N\"{a}gerl, \href{http://www.nature.com/nature/journal/v466/n7306/full/nature09259.html}{Nature {\bf 466,} 597 (2010).}

\bibitem{cold_bosons_many_spins_LL_velocities}
G. Pagano,	 M. Mancini,	 G. Cappellini,	 P. Lombardi,	 F. Sch\"{a}fer,	 H. Hu, X.-J. Liu,	 J. Catani,	 C. Sias,	 M. Inguscio, and L. Fallani, \href{http://www.nature.com/nphys/journal/v10/n3/full/nphys2878.html}{Nat. Phys. {\bf 10,} 198 (2014).}

\bibitem{LL_cold_atoms}
 A. Recati, P.O. Fedichev, W. Zwerger, and P. Zoller, \href{http://journals.aps.org/prl/abstract/10.1103/PhysRevLett.90.020401}{Phys. Rev. Lett. {\bf 90,} 020401 (2003)}; H.P. B\"{u}chler, G. Blatter, W. Zwerger, \href{http://journals.aps.org/prl/abstract/10.1103/PhysRevLett.90.130401}{Phys. Rev. Lett. {\bf 90,} 130401 (2003)}; B. Paredes and  J. I. Cirac, \href{http://journals.aps.org/prl/abstract/10.1103/PhysRevLett.90.150402}{Phys. Rev. Lett. {\bf 90,} 150402 (2003)}.

\bibitem{jc_lattice}
 M.J. Hartmann, F.G.S.L. Brandao, and M.B. Plenio, \href{http://www.nature.com/nphys/journal/v2/n12/full/nphys462.html}{Nat. Phys. {\bf 2,} 849 (2006)}; A.D. Greentree, C. Tahan, J.H. Cole and L.C.L. Hollenberg, \href{http://www.nature.com/nphys/journal/v2/n12/full/nphys466.html}{Nat. Phys. {\bf 2,} 856 (2006)}; D.G. Angelakis, M.F. Santos, and S. Bose, \href{http://journals.aps.org/pra/abstract/10.1103/PhysRevA.76.031805}{Phys. Rev. A {\bf 76,} 031805(R) (2007)}; K. Seo and L. Tian, \href{http://arxiv.org/abs/1408.2304}{arXiv:1408.2304 (2014)}.

\bibitem{jc_lattice_disorder}
A. Bermudez, M.A. Martin-Delgado, D. Porras, \href{http://iopscience.iop.org/1367-2630/12/12/123016/}{New J. Phys. {\bf 12,} 123016 (2010)}.

\bibitem{rabi_lattice}
M. Schir\'o, M. Bordyuh, B. Oztop, and H. E. T\"{u}reci, \href{http://prl.aps.org/abstract/PRL/v109/i5/e053601}{Phys. Rev. Lett. {\bf 109,} 053601 (2012)}; A. Kurcz, A. Bermudez, and J.J. Garcia-Ripoll, \href{http://journals.aps.org/prl/abstract/10.1103/PhysRevLett.112.180405}{Phys. Rev. Lett. {\bf 112,} 180405 (2014)}; P. Nevado and  D. Porras, \href{http://arxiv.org/abs/1406.5094}{arXiv:1406.5094 (2014)}.
 
 \bibitem{Lieb_robinson_spin_boson}
 J. J\"{u}nemann, A. Cadarso, D. P\'erez-Garc\'ia, A. Bermudez, and J. J. Garc\'ia-Ripoll, \href{http://journals.aps.org/prl/abstract/10.1103/PhysRevLett.111.230404}{Phys. Rev. Lett. {\bf 111,} 230404 (2013).}


 


\bibitem{XY_model}
E. H. Lieb, T. Schultz, and D. J. Mattis, \href{http://www.sciencedirect.com/science/article/pii/0003491661901154}{ Ann. Phys. {\bf 16,} 407
(1961)}.

\bibitem{jordan_wigner}
P. Jordan and E. Wigner, \href{http://link.springer.com/article/10.1007%2FBF01331938}{Z. Physik {\bf 47,} 631 (1928).}



\bibitem{bogoliubov_weakly_interacting}
N. N. Bogoliubov. \href{http://ufn.ru/dates/pdf/j_phys_ussr/j_phys_ussr_1947_11_1/3_bogolubov_j_phys_ussr_1947_11_1_23.pdf}{J. Phys. (USSR), {\bf 1,} 23 (1947)}.



\bibitem{lang_firsov}
I. G. Lang and Y. A. Firsov, Zh. Eksp. Teor. Fiz. {\bf 43,} 1843 (1962), (See also Y. A. Firsov, \href{http://link.springer.com/book/10.1007/978-1-4020-6348-0/page/1}{Small Polarons: Transport Phenomena}, in {\it Polarons in Advanced Materials}, Ed. A. S. Alexandrov, Springer Verlag, Bristol, 2007).

 \bibitem{supp_material}
See the Supplemental Material for technical details.

\bibitem{holstein}
T. Holstein, \href{http://www.sciencedirect.com/science/article/pii/0003491659900028}{Ann. Phys. {\bf 8,} 325 (1959)}; (ibid) \href{http://www.sciencedirect.com/science/article/pii/000349165990003X}{ {\bf 8,} 343 (1959)}.

\bibitem{comment_polaron}
For the dispersionless bosons of the Holstein model, a strong-coupling expansion that treats the electron hopping as a perturbation   $J_{\rm s}\ll \omega_{\rm b}, U$, leads to fermion-fermion interactions of strength $J_{\rm s}^2/\omega_{\rm b} U$ by a sort of super-exchange~\cite{holstein_strong_coupling}. This limit is not consistent with our treatment of the fermion-boson coupling  $U\ll J_{\rm s},J_{\rm b}$. Yet, the dispersive nature of our bosons allows for long-range interactions mediated by virtual boson exchange rather than by super-exchange.
 

\bibitem{holstein_strong_coupling}
J. E. Hirsch and E. Fradkin, \href{http://journals.aps.org/prb/abstract/10.1103/PhysRevB.27.4302}{Phys. Rev. B {\bf 27,} 4302 (1983)}.
 


 \bibitem{explicit_tunn}
 $\eta_\ell=\frac{u_0^2(1-\lambda^2)^{-\frac{3}{2}}}{(\omega_{\rm b}-U)^2}\left(1+\ee^{(\ii q_{\rm d}-\xi_0^{-1})d\ell}\left(1+\ell\sqrt{1-\lambda^2}\right)\right)$, for $\ell\geq 1$.
  
  \bibitem{explicit_int} $V_0=\frac{-2u_0^2}{(\omega_{\rm b}-U)}\frac{1}{\sqrt{1-\lambda^2}},$ and $\frac{d}{\xi_0}=-\log\left(\frac{1-\sqrt{1-\lambda^2}}{\lambda}\right).$


\bibitem{luther_peschel}
A. Luther and I. Peschel, \href{http://journals.aps.org/prb/abstract/10.1103/PhysRevB.12.3908}{Phys. Rev. B {\bf 12,} 3908 (1975).}


\bibitem{haldane_ll}
F. D. M. Haldane, \href{http://journals.aps.org/prl/abstract/10.1103/PhysRevLett.45.1358}{Phys. Rev. Lett. {\bf 45,} 1358 (1980)}.

\bibitem{dmrg}
S. R. White, \href{http://prl.aps.org/abstract/PRL/v69/i19/p2863_1}{Phys. Rev. Lett. {\bf 69,} 2863 (1992)}; {\it ibid} \href{http://journals.aps.org/prb/abstract/10.1103/PhysRevB.48.10345}{Phys. Rev. B {\bf 48,} 10345 (1993)}.



\bibitem{Heisenberg_dmrg}
K. A. Hallberg, P. Horsch, and G. Martinez, \href{http://journals.aps.org/prb/abstract/10.1103/PhysRevB.52.R719}{Phys. Rev. B {\bf 52,} R719 (1995)}.

\bibitem{xxz_nnn_dmrg}
C. Karrasch and  J. E. Moore, \href{http://journals.aps.org/prb/abstract/10.1103/PhysRevB.86.155156}{Phys. Rev. B {\bf 86,} 155156 (2012).}

\bibitem{capponi}
S. Capponi, D. Poilblanc, and T. Giamarchi, Phys. Rev. B {\bf 61,} 13410 (2000).

 
 \bibitem{const_bosonization}
J. von Delft and H. Schoeller, \href{http://onlinelibrary.wiley.com/doi/10.1002/(SICI)1521-3889(199811)7:4%3C225::AID-ANDP225%3E3.0.CO;2-L/abstract;jsessionid=0DFE1CC79E500BD82AB9BF6062FFA344.f01t01}{Ann. Phys. {\bf 7,} 225 (1998).}




\bibitem{explicit_cl}
$C_{\ell,\tilde{\ell}}=-\frac{u_0^2(1-\lambda^2)^{-\frac{3}{2}}}{4(\omega_{\rm b}-U)^2}\ee^{\ii q_{\rm d}d\tilde{\ell}}\left(1+|\ell-\tilde{\ell}|\sqrt{1-\lambda^2}\right)$.

\bibitem{itebd}
R. Orus and G. Vidal, \href{http://prb.aps.org/abstract/PRB/v78/i15/e155117}{Phys. Rev. B {\bf 78,} 155117 (2008)}.

\bibitem{wineland_review}
{\it See} D. J. Wineland, C. Monroe, W. M. Itano, D. Leibfried, B. E.
King, and D. M. Meekhof, \href{http://arxiv.org/abs/quant-ph/9710025}{J. Res. Natl. I. St. Tech. {\bf 103,} 259 (1998)}, {\it and references therein.}

\bibitem{devoret_review}
{\it See} M. H. Devoret and  J.M. Martinis, \href{http://link.springer.com/article/10.1007/s11128-004-3101-5}{Quant. Inf.Processing, Volume {\bf 3,} 163 (2004)}, {\it and references therein.} 








\end{references}
\end{document}